\documentclass[prd,twocolumn]{revtex4}

\usepackage{amsmath,amsthm,amssymb}
\usepackage{graphics,graphicx}
\usepackage{epsfig}
\usepackage{color}
\makeatletter
\@addtoreset{equation}{section}
\makeatother
\begin{document}
\preprint{UVIC-TH-09-01}
%%%%%%%%%%%%%%%%%%%%%%%%%%%%%%%%%%%%%%%%%%%%%
\title{\Large An alternative approach to tachyon cosmology}
\author{John Ward \footnote{email:jwa@uvic.ca}}

\affiliation{
  Department of Physics and Astronomy, University of Victoria,
  Victoria, BC, V8P 5C2, Canada
  }

\date{December 2008}
\begin{abstract}
In this note we propose the use of an alternate action for the open string tachyon on a non-BPS $D3$-brane. At the classical level this action
is precisely equivalent to the more commonly used DBI action, but involves an additional coupling to dynamical world-volume gravity. We find
that, for a FRW metric, exponential expansion occurs provided that the cosmological constant is positive. For anti de-Sitter solutions we find a periodically
bouncing universe, and there are no accelerating trajectories for a theory with no cosmological constant. 
In specific cases the acceleration is terminated by the condensation of the open string tachyon, leading to a canonical inflationary trajectory at late times.
\end{abstract}
\maketitle
%%%%%%%%%%%%%%%%%%%%%%%%%%%%%%%%%%%%%%%%%%%%%
\section{Introduction}
Developing realistic inflationary models of cosmic inflation within string theory has been a dominant area of research in recent years.
New and improved models are now proposed on a regular basis, and it is fair to say that some of the predictions have led
to exciting insights in cosmology \cite{Spergel:2006hy, Burgess:2007pz}.

One of the first, and simplest, models of inflation within a string theory context was based upon the condensation of the open string tachyon that
exists on the world-volume of non-BPS $Dp$-branes in the type II theory \cite{Sen:2002nu,Sen:2004nf}. 
In order to have a consistent coupling with $3+1$ dimensional gravity, the 
large transverse dimensions had to be compactified. Initially this was done by hand, but with breakthroughs in our understanding of moduli stabilisation,
this was extended to more realistic scenarios using warped backgrounds. 
It was found that tachyon inflation could occur, provided that the fluxes generating the background were extremely finely tuned to ensure that the Hubble
friction allowed for sufficient inflation to occur, since the mass (squared) of the tachyon field was generically too large and needed to
be reduced through the effects of warping \cite{Raeymaekers:2004cu}. In order to save the model, many proposals were put forward - including assisted tachyonic 
inflation \cite{Singh:2006yy}, which required
$N$ non-BPS branes to exist in close proximity; each having a condensing tachyon. Again one found that upon comparison with data from WMAP \cite{Spergel:2006hy}, that the 
required value of $N$ was extremely large and most-likely unphysical.
Whilst fully realistic models are complicated and have yet to be constructed, the simplicity of the tachyon model
as an example of a Chaplygin gas suggested that it may still find some use as a model of dark energy. This area is currently under intense
scrutiny, see eg. \cite{Copeland:2006wr}. 

Perhaps another pressing problem relates to the origin of the gravitational coupling. Compactifying upon a six-manifold does indeed
lead to a minimally coupled gravity theory, however this is still a two-step process in some sense. Firstly one aims to stabilise the geometric
moduli of the compact space, then one considers how $D$-branes couple to the resulting $3+1$ dimensional gravity. It would be preferable to
include the interactions of the open string moduli with the closed string sector prior to this step.
There has been some progress along this direction, but it is clearly a technically complex problem.

In this note we wish to propose an alternative cosmological model of tachyon inflation, using a $D$-brane action that is classically equivalent
to the standard DBI one. This is a generalisation of the models proposed in \cite{AbouZeid:2000nf, Polyakov:1981rd} to describe $D$-brane systems. 
What is intriguing is that there
is a coupling to a world-volume metric which, when made dynamical, induces a non-minimal coupling to $3+1$ dimensional gravity. Thus the cosmology
of a world-volume observer in the theory is similar in spirit to bulk theories such as Mirage Cosmology \cite{Kehagias:1999vr}.
%%%%%%%%%%%%%%%%%%%%%%%%%%%%%%%%%%%%%%%%%%%%%%
\section{Non-minimally coupled tachyon theory}
There has been a large amount of work on the description of non-BPS $Dp$-branes in type II string theory, motivated by the Sen conjectures and
developments in boundary string theory \cite{Sen:2004nf}. Whilst the BSFT (Boundary String Field Theory) 
description of tachyon condensation is relatively robust, the effective theory has often
been described by the non-BPS action discussed in \cite{Sen:1999md}. This action takes the usual 'square root' form, familiar to the the DBI action commonly
used in the description of BPS $D$-branes (neglecting the Chern-Simons coupling)
\begin{equation}\label{eq:dbi}
S = -T_p \int d^{p+1}\sigma V(T) e^{-\phi} \sqrt{-\rm{det}(\hat{G}_{ab}+\lambda \partial_a T \partial_b T)}
\end{equation} 
where $\hat{G}_{ab}$ is the pullback of the bulk metric to the world-volume and $\lambda$ is the inverse $F$-string tension.
However there is an alternate description which removes the square root term, and allows for a
more concrete study of open string modes. This formalism depends upon the existence of an auxiliary metric $\gamma_{ab}$ which lives on the
world-volume of the $D$-branes. For a generic auxiliary metric, one finds that the resulting action is linear in derivatives and the gauge field strength.
The specific case of interest for our purpose is when the metric $\gamma_{ab}$ satisfies $g_{ab}=\gamma_{(ab)}$, $\gamma_{[ab]}=0$ 
in which case the metric is (again) linear
in scalar derivatives, but quadratic in the gauge field strength. Thus this is far more amenable regarding quantisation of open string states.

Using this formalism, we propose the generalisation of the BPS states to include non-BPS $Dp$-branes using a new action of the form \cite{AbouZeid:2000nf}
\begin{eqnarray}\label{eq:newaction}
S &=& -T_p '' \int d^{p+1} \sigma V(T) e^{-\phi}(-\hat{G})^{1/4}(-g)^{1/4} \times \nonumber \\ 
& &(g^{ab}\hat{G}_{ab}-(p-3)\Lambda)
\end{eqnarray}
where $\hat{G}$ is now the linear combination of the induced metric and the (canonical) open string tachyon coupling
\begin{equation}
\hat{G}_{ab} = \partial_a X^{A} \partial_b X^{B} G_{AB}+ \lambda \partial_a T \partial_b T
\end{equation}
and $V(T)$ is the usual tachyonic potential. We have also set the gauge field to zero. The tension of the brane is given by
\begin{equation}
T_p'' = \frac{T_p}{4}\Lambda^{(p-3)/4}
\end{equation}
where $\Lambda$ represents a world-volume cosmological constant term, and the open string sector has its indices raised and lowered with the world-volume metric $g_{ab}$. 
%This is clearly an example of a bi-metric theory, which has found recent application in the relativity literature.
One can check that the equation of motion for the world-volume metric yields the relation
\begin{equation}
g_{ab} = \Lambda^{-1} \hat{G}_{ab}
\end{equation}
and upon substituting this back into (\ref{eq:newaction}) one recovers the action (\ref{eq:dbi}). Thus, at least at the classical level, these
two actions are equivalent. In what follows we will restrict ourselves to the cosmological case of interest, which is the $p=3$ solution. Note that
this the cosmological constant term vanishes from the action without the need to set $\Lambda=0$. 

To discuss the role of the tachyon in cosmology, we need to couple the system to gravity. Typically in the case of the DBI form of the action, this requires
us to embed the action into a consistent compactification of string theory \cite{{Burgess:2007pz}}. 
For a certain class of compactifications, this yields $3+1$ dimensional Einstein
gravity and we are allowed to write the low-energy effective action as a minimally coupled theory
\begin{equation}
S = \int d^4 \sigma \sqrt{-g}\left( M_p^2 R + S_{DBI}\right)
\end{equation}
where $M_p$ is the (reduced) Planck mass which is fixed by the volume of the compact manifold, and $g$ is the determinant
of the Einstein-frame metric, which we anticipate will be of the FRW (Friedmann-Robertson-Walker) form. 
On the other hand, the action (\ref{eq:newaction}) already contains
a coupling to a world-volume metric which we will promote to a dynamical field.
All that is required is that we include the relevant kinetic terms for the graviton,
and this leads us to the following form of the effective action
\begin{equation}
S = \int d^4 \sigma \sqrt{-g} \left(a \lambda R + a \Lambda_2 + S_{new}\right)
\end{equation}
where both $a$ and $\lambda$ are (positive) coupling constants and $\Lambda_2$ is an additional cosmological constant term that may be present. Note that the 
strength of gravity is now set by the combination $a\lambda$ which plays the role of the Planck scale. However because we now promote the world-volume metric
$g_{ab}$ to be dynamical, this means that we have a complicated, non-minimally coupled theory similar to those discussed in \cite{Moffat:1997cc}.

To analyse the cosmological dynamics of the theory we choose to embed the tachyon action into a warped ten-dimensional space, consistent with
known $Dq$-brane solutions of type II supergravity. Typically the bulk metric will be of the following form \cite{Burgess:2007pz}
\begin{equation}
ds^2 = h^{-1/2}(y) \eta_{\mu \nu} dx^\mu dx^\nu+ h^{1/2}(y)g_{mn}dy^m dy^n
\end{equation}
where $h(y)$ is a harmonic function in the $(9-q)$ transverse directions to the $Dq$-branes and $g_{mn}$ is the metric in these directions.
Note that the space-time should be (approximately) asymptotically Minkowski.
For simplicity let us assume that $q \ge 3$ since this means that the term contributing to the induced metric (in the limit of a rigid brane) will be
$\hat{G}_{ab} = h^{-1/2}\eta_{ab}$ where $h$ is a constant. Physically this corresponds to our $D3$-brane being localised at some point
in the bulk space-time, which may well be an atypical situation, but serves as a useful toy model. 
We will also consider a homogeneous tachyon field $T(x^0)$ and study the dynamical interplay between the tachyon and the metric. The tachyon
potential itself will be kept general, although we note that it must vanish as $T \to \pm \infty$.

Cosmological solutions typically demand an isotropic and homogeneous metric, which we will take to be of the FRW form in the $p+1$ directions.
\begin{equation}
ds^2 = -dt^2 + b(t)^2 d\vec{x}^2
\end{equation}
where $b(t)$ will play the role of the scale factor of the universe on the brane.
Employing the use of static gauge we see that the independent $(p+1)$-dimensional Einstein equations become, using $X=(1-h^{1/2}\lambda \dot{T}^2)$
\begin{eqnarray}
H^2 &=& \frac{T_p'' V(T)e^{-\phi}}{2 \lambda_1 a b^{p/4} h^{(p+1)/8}} \times \nonumber \\
& & \frac{X^{1/4}}{(p(p-3)+2)} \left(\frac{3X}{h^{1/2}}-\frac{p}{b^2 h^{1/2}}-(p-3)\Lambda_1 \right) \nonumber \\
&+& \frac{\Lambda_2}{a(p(p-3)+2)} \nonumber \\
\frac{\ddot{b}}{b} &=& - \frac{T_p'' V(T)e^{-\phi}}{4 h^{(p+1)/8}\lambda_1 a (p-2)} \times \nonumber \\
& & \frac{X^{1/4}}{b^{(p+8)/4}} \left(b^2 \Lambda_1(p-3)+\frac{(p-4)}{h^{1/2}}+\frac{b^2X}{h^{1/2}} \right) \nonumber \\
&+& \frac{\Lambda_2}{2a(p-2)} - \frac{\alpha_1}{(p-2)}
\end{eqnarray}
where $\alpha_1$ is defined as $\alpha_1 = \frac{p}{2}(p-5)+3$, $\lambda$ is the usual open string coupling and $H$ is the usual Hubble
parameter defined as $H=\dot{b}/b$. We have kept
the open string coupling explicit in these equations, although we know that the dilaton is generally solved in terms of the harmonic function.
Note that the Einstein equation does not admit a vacuum solution unless the $Dp$-brane is tensionless.

Specialising to the $D3$-brane case relevant for cosmology, we see that the above equations can be combined to yield
\begin{equation}\label{eq:gravitydynamics}
H^2+\frac{3 \ddot{b}}{b}- \frac{2 \Lambda_2}{a} = 0
\end{equation}
irrespective of the position or velocity of the tachyon field. The decoupling of gravity from the tachyonic modes occurs because of the unusual algebraic 
structure of the tachyon action in $3+1$ dimensions. This allows us to eliminate the tachyon dynamics from the problem, leaving us with a modfied gravitational
theory. The tachyon field equation can then be imposed as a constraint on the gravitational solution.
Note that this is true because of our initial assumption that the brane has no radial dynamics, which ensures that the tachyon sector is not as sensitive 
to the particular bulk theory in which it is embedded. This is vastly different from the situation that occurs in standard tachyon
inflation, which requires precise tuning of the warp factors to ensure that inflation occurs \cite{Raeymaekers:2004cu}.
We expect equation (\ref{eq:gravitydynamics}) to break down in two places. Firstly when the tachyon
has rolled to the point where the potential vanishes - since this effectively makes the brane tensionless and we recover the usual vacuum Einstein solution. The
second place where we expect this expression to be valid is when the tachyon rolls at the (warped) relativistic limit. In studies with the DBI form
of the tachyon action, these two conditions typically arise at the same place.

The general solution to the above expression depends upon the sign of the cosmological constant. Let us initially consider an Anti de-Sitter solution,
whereby we see that
\begin{equation}
b(t) = b_0 \cos \left( \sqrt{\frac{2|\Lambda_2|}{a}}\left(\frac{3t}{2}-\frac{\pi}{4}\sqrt{\frac{2a}{|\Lambda_2|}} \right) \right)^{3/4}
\end{equation}
and where $b_0$ is a constant of integration which we expect to be set by the tachyon condensate. We have imposed boundary conditions such that the 
scale factor vanishes at $t=0$. The profile for the scale factor implies that this is a bouncing universe, which collapses to zero size at a time
\begin{equation}
t_{crit} = \frac{m \pi}{3} \left(1+\sqrt{\frac{a}{2m^2|\Lambda_2|}} \right)
\end{equation}
for some $m \in \mathbb{Z}$. The maximum size occurs is set by the scale $b_0$ and occurs at
\begin{equation}
t_{max} = \frac{\Pi}{6} \sqrt{\frac{2 a }{|\Lambda_2|}}
\end{equation}
in the fundamental domain. Physically both conditions imply $a >> |\Lambda_2|$ as a constraint on the cosmological constant.
This solution is interesting because the 
scale factor never accelerates, and therefore there is no sense that this is a standard inflationary scenario. To fully understand the dynamics of the universe
we must also consider the condensation of the tachyon. Since $V(T)$ is a decreasing function, 
we anticipate that this modulates the value of $b_0$ and therefore the overall size of the universe decreases with each bounce. However since $b(t)$ is
no longer a monotonic function this will lead to a complicated profile for the tachyon condensate.

After some time $t_f$ we will find $V(T) \to 0$ and the corresponding solution for 
the scale factor becomes constant, modulated by a phase of the form 
\begin{equation}
b \sim b_0(t_f) \exp \left(it_{*} \sqrt{\frac{|\Lambda_2|}{2a}} \right)
\end{equation}
where $0 \le t_f \le t_{*}$. Thus the bouncing universe phase naturally becomes of constant size (set by $b_0$ at the end of tachyon condensation), albeit it
with a (non-physical) complex phase

If there is a de-Sitter solution on the other hand, we see that the scale factor is a complex hyperbolic function with a phase again set by the ratio
of $a/\Lambda_2$. Ignoring the complex phase, we can write the \emph{real} part of the solution as
\begin{equation}\label{eq:dsscalefactor}
b(t) = b_0 \cos \left(3 \pi \sqrt{\frac{2a}{\Lambda_2}} \right)^{3/4} \cosh \left(\sqrt{\frac{2 \Lambda_2}{a}}\frac{t}{4} \right)^{3/4}
\end{equation}
where we have again assumed the existence of a singularity. From this equation  we see that there is a bound on $a/\Lambda_2$ such that
\begin{equation}
\left( \frac{a}{\Lambda_2}\right)  < \frac{1}{2} \left(\frac{1}{6} \right)^2
\end{equation}
in the fundamental domain - suggesting that the cosmological constant needs to be tuned to  $\sim \mathcal{O}(10^2)$ times the gravitational scale $a$. 
There is a secondary branch of solutions leading to a bouncing universe, with scale factor 
\begin{equation}
b(t) = b_0 \cosh \left(\sqrt{\frac{2 \Lambda_2}{a}}\frac{2t}{3} \right)^{3/4}
\end{equation}
which describes a smooth transition between a contracting/expanding phase, where the universe collapses down to a minimal size $b_0$ - before undergoing accelerated 
expansion provided that $(\Lambda_2/a)$ is not sufficiently suppressed. If $b_0$ is sufficiently small
then this implies we evade the singularity problem, and also find an inflationary solution due to the algebraic structure of the hyperbolic cosine.
This solution has the additional bonus in that the scale factor is purely real. However in this case we are unable to constrain the value of the
cosmological constant.
At late times we know that the tachyon condensate forces the 
scale factor to be exponential, however since $\Lambda_2$ is positive definite, we recover the canonical inflationary trajectory after tachyon
condensation. Thus any inflation occurring whilst the tachyon condenses can be regarded as a 'pre-inflationary' epoch. The net effect is to reduce the overall
number of e-foldings required in the standard inflationary phase. This is true for both models discussed here.

If we now study the case where $\Lambda_2=0$ we find that the solution is simply given by a power law
\begin{equation}
b(t) = b_0 \left(\frac{t}{t_f} \right)^{3/4}
\end{equation}
where $t_f$ marks the end of the initial phase.
However this is not an inflationary solution because we see that $\ddot{b}$ is negative definite because the power of $t$ is less than unit. 
Moreover once the tachyon condenses we see that $H^2=0$, implying that $b(t)$ is constant as in the case of Anti de-Sitter space.

As this is a coupled non-linear system, we must also consider the field equation for the tachyon condensate - which acts as a constraining condition
\begin{eqnarray}
0 &= & V'(T) X^{1/4}(X+3 b^{-2}) \\ 
&+& \frac{\partial}{\partial t} \left(\frac{5 V(T)h^{1/2}X^{1/4} \lambda \dot{T}}{4}\left\lbrack 1+\frac{3}{5 b^2 X} \right\rbrack \right) \nonumber
\end{eqnarray}
where a prime denotes a derivative with respect to $T$. Dropping all acceleration terms the equation of motion reduces to 
\begin{eqnarray}
0 &\sim & 2X V'(T) \left(4Xb^2(Xb^2+3)+\lambda h^{1/2} \dot{T}^2(5Xb^2+3) \right) \nonumber \\
&-& 3 \lambda h^{1/2}V(T)\dot{T}H .
\end{eqnarray}
Unfortunately even the simplest case is non-analytic. What is clear is that at $t=0$, the velocity must be zero. At late times one sees that $X \to 0$ which
dominates the scale factor terms and ensures that the equation of motion is satisfied. However the profile for the condensate at intermediate times
cannot be determined. One sees that provided $T=\dot{T}=0$ at $t=0$, even the dS bounce is trivially satisfied.
%%%%%%%%%%%%%%%%%%%%%%%%%%%%%%%%%%%%%%%%%%%%%%
\section{Discussion}
We have seen that the new proposed action admits FRW-type cosmology provided that the cosmological constant is non-zero.
In the AdS case we find a bouncing
universe which collapses to zero size, although there is never any inflation. In the dS case we find more complicated behaviour. There is a solution
emerging from a singularity, and also a collapsing/expanding solution where the universe has a minimal size $b_0$. In both cases we find inflationary
solutions due to the hyperbolic cosine. The issue here concerns the ending of inflation, which is beyond the scope of this note, but we 
note that this is likely to occur if the brane has non-zero velocity in the bulk space-time due to a modification of the Einstein equations.
The AdS solution is interesting since it yields a bouncing cosmology which naturally terminates after tachyon condensation. Because the tachyon
potential is a decreasing function, we expect the amplitude of the bounce to decrease with time until it becomes constant, albeit modulated by a
phase set by the cosmological constant. The collapsing solution appears to be the most natural in the dS framework, allowing us to avoid the initial singularity 
problem whilst still connecting to late time inflation. 

In principle this is a coupled system, since the tachyon equation of motion depends explicitly upon the evolution of the world-volume metric, however we 
anticipate that if the solution is tuned to ensure that the tachyon stays near the origin - then sufficient inflation will occur. Unfortunately this (additional) 
fine-tuning is a requirement of our theory, and it is hard to see how it can be avoided.

Since our model exists on a brane inside the large ten-dimensional universe, one can speculate that the $D3$-brane is a remnant of a brane/anti-brane
annihilation cascade. Thus the universe began in an initially (hot) phase consisting of $D9-\bar{D}9$-pairs which eventually annihilated as the universe cooled. The
annihilation mechanism creates daughter branes of various co-dimension, until we cascade down to a $\bar{D}3$-brane. It is tempting to speculate that the 
brane version of the Brandenberger-Vafa mechanism may hold here \cite{Brandenberger:1988aj}, since the brane is uncharged with respect to the $RR$-form fields.

Whilst this is an interesting idea, it is only a toy model of inflation. As a world-volume theory the $D$-brane only carries a $U(1)$ gauge theory,
and therefore cannot describe the standard model \cite{Polchinski:1996na}. 
If there were $N$ branes with tachyon condensation occurring on one of them, the gauge group is enhanced
from $U(1)^N \to U(N)$ in an appropriate limit, however there are no chiral fermion states unless the branes intersect at an angle \cite{Myers:2003bw}.
Interestingly the trigonometric dependence of the scale factor remains unmodified even if we take the non-BPS brane to intersect with the background branes, 
provided (of course) that it remains fixed in the transverse space. This would then give rise to chiral matter on the background branes.
Additionally there is no inherent scale which fixes the magnitude of world-volume cosmological constant, although we see in the dS case that there are some 
weak bounds. Needless to say,
there is significant work required to put this toy model onto firmer footing. However there are many potentially interesting applications, such
as finding black-hole solutions on the brane and perhaps even a 'little' version of the AdS/CFT duality - should AdS solutions be allowed in such a system. We hope to 
return to such issues in a future publication.
\section*{Acknowledgements}
We thank Narit Pidokrajt for his comments.
%%%%%%%%%%%%%%%%%%%%%%%%%%%%%%%%%%%%%%%%%%%%%%
%\subsection{Tachyonic equation of motion}
%Because of the unusual coupling to gravity, the energy density of the tachyonic condensate is not conserved. This complicates the resulting
%analysis of the tachyon dynamics since we cannot write the solution in terms of a conserved charge.
%Denoting $X=(1-h^{1/2}\lambda \dot{T}^2)$ for simplicity, we find that the resulting equation of motion for the tachyon condensate
%\begin{equation}
%V'(T) X^{1/4}(X+3 b^{-2}) + \frac{\partial}{\partial t} \left(\frac{5 V(T)h^{1/2}X^{1/4} \lambda \dot{T}}{4}\left(1+\frac{3}{5 b^2 X} \right) \right) =0
%\end{equation}
%where a prime denotes a derivative with respect to $T$. Dropping all acceleration terms, the equation of motion reduces to the simpler expression
%\begin{equation}
%2X V'(T) \left(4Xb^2(Xb^2+3)+\lambda h^{1/2} \dot{T}^2(5Xb^2+3) \right)-3 \lambda h^{1/2}V(T)\dot{T}H \sim 0
%\end{equation}
%%%%%%%%%%%%%%%%%%%%%%%%%%%

\end{document}